\documentclass[runningheads]{llncs}
\usepackage{hyperref}

\usepackage{graphicx}

\usepackage{todonotes}

\usepackage[inline]{enumitem}
\usepackage{caption}
\usepackage{subcaption} %
\usepackage{algorithm}
\usepackage[noend]{algpseudocode}
\usepackage{etoolbox}
\usepackage{lipsum}
\usepackage{comment}
\usepackage{amsmath}
\usepackage{wrapfig}
\usepackage{amssymb}

\makeatletter

\newcommand*{\algrule}[1][\algorithmicindent]{%
  \hspace*{.2em}%
  \vrule height .75\baselineskip depth .25\baselineskip
  \hspace*{\dimexpr#1-.2em-.4pt}%
}

\newcommand{\StatePar}[1]{%
  \State\parbox[t]{\dimexpr\linewidth-\ALG@thistlm}{\strut #1\strut}%
}
\renewcommand{\ALG@beginalgorithmic}{\offinterlineskip}%

\newcount\ALG@printindent@tempcnta
\def\ALG@printindent{%
  \ifnum \theALG@nested > 0%
    \ifx\ALG@text\ALG@x@notext%
    \else
      \unskip
      \ALG@printindent@tempcnta=1
      \loop
        \algrule[\csname ALG@ind@\the\ALG@printindent@tempcnta\endcsname]%
        \advance \ALG@printindent@tempcnta 1
        \ifnum \ALG@printindent@tempcnta<\numexpr\theALG@nested+1\relax
      \repeat
        \fi
    \fi
}
\patchcmd{\ALG@doentity}{\noindent\hskip\ALG@tlm}{\ALG@printindent}{}{\errmessage{failed to patch}}
\makeatother

\algrenewcommand\algorithmicend{\strut\textbf{end}}
\algrenewcommand\algorithmicdo{\strut\textbf{do}}
\algrenewcommand\algorithmicwhile{\strut\textbf{while}}
\algrenewcommand\algorithmicfor{\strut\textbf{for}}
\algrenewcommand\algorithmicforall{\strut\textbf{for all}}
\algrenewcommand\algorithmicloop{\strut\textbf{loop}}
\algrenewcommand\algorithmicrepeat{\strut\textbf{repeat}}
\algrenewcommand\algorithmicuntil{\strut\textbf{until}}
\algrenewcommand\algorithmicprocedure{\strut\textbf{procedure}}
\algrenewcommand\algorithmicfunction{\strut\textbf{function}}
\algrenewcommand\algorithmicif{\strut\textbf{if}}
\algrenewcommand\algorithmicthen{\strut\textbf{then}}
\algrenewcommand\algorithmicelse{\strut\textbf{else}}

\algrenewcommand\algorithmicrequire{\strut\textbf{Input:}}
\algrenewcommand\algorithmicensure{\strut\textbf{Output:}}

\let\oldState\State
\renewcommand{\State}{\oldState\strut}

\begin{document}
\title{SymED: Adaptive and Online Symbolic Representation of Data on the Edge}

\author{Daniel Hofstätter\inst{1} \and Shashikant Ilager\inst{1} \and Ivan Lujic\inst{2} \and Ivona Brandic\inst{1}}

\authorrunning{D. Hofstätter et al.}

\institute{ 
Vienna University of Technology, Austria\\
\email{\{daniel.hofstaetter, shashikant.ilager, ivona.brandic\}@tuwien.ac.at}
\and Ericsson Nikola Tesla, Croatia\\
\email{ivan.lujic@ericsson.com}
}
\maketitle              %

\begin{abstract}
The edge computing paradigm helps handle the Internet of Things (IoT) generated data in proximity to its source. Challenges occur in transferring, storing, and processing this rapidly growing amount of data on resource-constrained edge devices. Symbolic Representation (SR) algorithms are promising solutions to reduce the data size by converting actual raw data into symbols. Also, they allow data analytics (e.g., anomaly detection and trend prediction) directly on symbols, benefiting large classes of edge applications. However, existing SR algorithms are centralized in design and work offline with batch data, which is infeasible for real-time cases. We propose SymED - Symbolic Edge Data representation method, i.e., an online, adaptive, and distributed approach for symbolic representation of data on edge. SymED is based on the Adaptive Brownian Bridge-based Aggregation (ABBA), where we assume low-powered IoT devices do initial data compression (senders) and the more robust edge devices do the symbolic conversion (receivers). We evaluate SymED by measuring compression performance, reconstruction accuracy through Dynamic Time Warping (DTW) distance, and computational latency. The results show that SymED is able to (i) reduce the raw data with an average compression rate of $9.5\%$; (ii) keep a low reconstruction error of $13.25$ in the DTW space; (iii) simultaneously provide real-time adaptability for online streaming IoT data at typical latencies of 42ms per symbol, reducing the overall network traffic.

\end{abstract}
\section{Introduction}
\label{sec:introduction}

The Internet of Things (IoT) enables various physical devices to embed with sensors and actuators to exchange data with smart systems over the Internet. Rapid growing IoT data are traditionally transmitted to a centralized cloud to derive insights for smart applications. However, this remote cloud-centric approach does not satisfy time-critical IoT application requirements~\cite{satyanarayanan2017emergence,trivedi2020sharing} and can create network congestion~\cite{ranjan2014streaming}. Consequently, edge computing mitigates these issues by delivering computing, storage, and network resources at the network edge. 
\\ \indent  Edge nodes are highly distributed resource-limited devices deployed in the proximity of IoT data sources to deliver time-critical processing~\cite{satyanarayanan2017emergence}.
Unlike the cloud, edge nodes have limited computation and storage resources. Therefore, it becomes crucial for edge nodes to cope with the velocity and growing volume of data generated and support applications within their resource constraints.
Several efforts have been made to reduce network traffic and improve data storage using edge data processing techniques. In~\cite{papageorgiou2015real}, authors target edge data reduction focusing on IoT data and adapting a posteriori data reduction techniques to data streams. Nevertheless, this approach does not consider the impact of reduced data on data analytics tasks. Consequently, Symbolic Representation (SR) techniques are promising alternative methods to reduce the data size while maintaining partial semantics of the data~\cite{lin2007experiencing}. 
\\ \indent The SR helps reduce the dimension and volume of time series, enabling efficient edge data storage management. The raw data in SR are segmented and represented with symbols that can be reconstructed to their original dimension. Unlike common raw data compression methods, the symbolically converted data in SR can help to directly perform data mining tasks such as pattern matching, substring search, motif discovery, and time series prediction, which are commonly used techniques in IoT applications \cite{elsworth2020abba}. However, the state-of-the-art SR algorithms are designed for centralized batch processing systems and perform an offline conversion, where often fixed parameters (e.g., window and alphabet size) are needed, making them infeasible for streaming data in modern IoT systems.
\\ \indent We propose \textit{SymED} (\underline{Sym}bolic \underline{E}dge \underline{D}ata representation) approach, i.e., an online distributed and adaptive SR method suiting edge data storage management and transmission. SymED is based on the Adaptive Brownian bridge-based symbolic aggregation (ABBA) algorithm, due to its adaptiveness in window and alphabet size. We decompose the algorithm into distributed manner with two main components: sender and receiver.
We also incorporate online normalization and clustering for adaptation to streaming data and symbol conversion.
Furthermore, SymED allows us to adaptively adjust the reconstruction error and bandwidth usage between sender and receiver depending on hyperparameter configurations.
The main contributions include \textbf{(i)} a symbolic representation approach for IoT sensor-based time series, investigating the benefits of edge storage and transmission bandwidth scarcity;
\textbf{(ii)} an online symbolic representation algorithm for real-time symbol generations in edge environments; \textbf{(iii)} an empirical evaluation of the proposed solution on real-world data sets, showing different performance profiles and achieving raw data compression of 9.5\% on average while minimizing reconstruction error.

\section{Motivation and Background}
\label{sec:background}

\textbf{Need for Symbolic Representation on Edge:} SR methods are promising solutions that allow analytic tasks to be performed directly on reduced data and enable the reconstruction of original data with minimal error. Existing symbolic representation algorithms have limited applicability for edge due to the following design requirements: (1)~\textbf{Online}: Compression should be continuous and immediate (i.e., stream-based).
(2) \textbf{Adaptive:} A SR algorithm should be adaptive, allowing flexible compression and reconstruction performance based on application and resource constraints. (3) \textbf{Distributed:} A SR should be distributed in edge as IoT sensors themselves do not have enough computational/network capabilities.
Existing SR algorithms assume apriori availability of batch data and work offline in a centralized manner. 
\begin{figure*}[!t]
	\includegraphics[width=\linewidth]{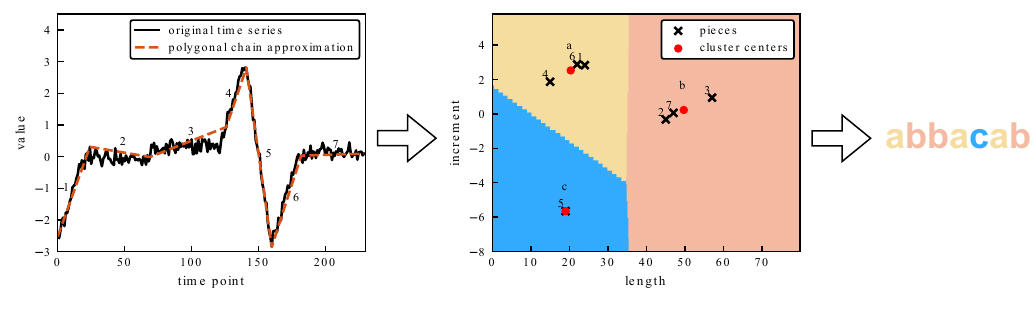}
 \caption{Illustration of ABBA~\cite{elsworth2020abba}.  (i) Creating polygonal chain of linear pieces (left side).  (ii) Clustering pieces (middle). (iii) Symbolizing (right side), i.e., \textit{abbacab}.}
	\label{fig:edgeArch}
\end{figure*}
\\
\indent \textbf{Symbolic Representation for Time Series Data:} 
A SR algorithm transforms time series into a string using finite alphabet size. Let us consider a time series $T =[t_0, t_1,...,t_N] \in \mathbb{R}^{N+1}$ converted into a symbolic representation $S = [s_1, s_2,...,s_n] \in \mathbb{A}^n $, with symbols $s_i$ coming from an alphabet of $k$ symbols $\mathbb{A} =\{a_1, a_2,...,a_k\}$~\cite{elsworth2020abba}. The sequence $S$ should be of considerably lower dimension than the original time series $T$, that is $n << N$, and it should only use a small number of meaningful symbols, that is $k << n$. The symbolic representation must also allow a reconstruction, with (i) a minimal and controllable error, and (ii)  a shape suitably close to the original time series data. 
\\ 
\indent \textbf{ Adaptive Brownian Bridge-based Aggregation (ABBA):} Our SymED is based on ABBA symbolic method and adapted for edge environments. Figure~\ref{fig:edgeArch} shows an example of ABBA symbolic conversion, with the black line on the left side as the original data, and the symbolically represented data on the rightmost side. 
ABBA adaptively finds linear pieces (7 red dashed lines on the left), where similar pieces are clustered together based on their length and increment values (middle), and each cluster is mapped to a symbol from the alphabet, resulting in a string (right). A tolerance hyperparameter $tol$ sets boundaries for the allowed reconstruction error, where a lower value results in a lower reconstruction error, but also a lower compression rate with more symbols.
In this example, $230$ data points are converted to a  word of  just $7$ symbols (rightmost part of Figure~\ref{fig:edgeArch}). A similar inverse approach will be applied during the  reconstruction of the data. However, many challenges arise when using such algorithms for online and resource-constrained edge environments, which we address in this work.

\section{SymED: Symbolic Edge Data representation} \label{sec:conversion}

\begin{wrapfigure}{r}{0.45\textwidth}
 \includegraphics[width=0.44\textwidth]{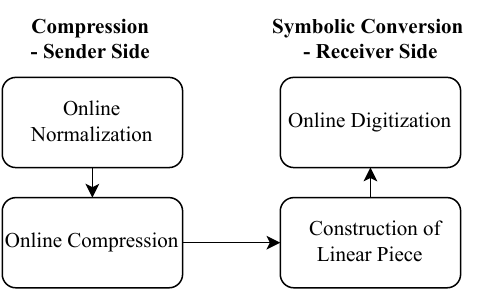}
              \caption{SymED Components.}
\label{fig:components}
\end{wrapfigure}

We present SymED as an online and adaptive symbolic representation method for streaming IoT data. Figure~\ref{fig:components} shows the SymED components. Our goal is to enable distributed symbolic representation where raw data communication and storage usage are limited in IoT-edge environments.
A sender (IoT node) normalizes and compresses all incoming data. A receiver (edge node) collects transmitted data to 
\begin{enumerate*}[label=(\roman*)]\item construct linear pieces (line segments), \item converts them to symbols in the digitization phase, and \item optionally reconstructs pieces or symbols again.
\end{enumerate*}

\subsection{Sender Side - Compression}
The sender compresses data stream $T=[t_0, t_1, ...., t_N]$ for each new data point $t_j \in T$ step-wise. Our compression technique, leverages the existing method~\cite{elsworth2020abba} to an online setting, with additional online normalization, shown in Algorithm~\ref{alg:SymED sender online compression}.
The sender collects and normalizes data stream points $[t_0, t_1, ... , t_m]$ ($m<<N$), and fits them to a linear line. After transmitting only the end point of this line to the receiver, the whole process repeats. 
\\
\indent \textbf{Online Normalization:}
Using normalized data is mandatory for a uniform conversion performance, as data can arrive with arbitrary scaling. A popular normalization technique is the Min-Max-Normalization~\cite{attig2011problem,gupta2019adaptive}. 
We use Z-Score-Normalization (standardization) that provides scaling data with zero mean and unit variance. Standardization in an online setting is used for, e.g., improving batch normalization in continual learning~\cite{pham2021continual}.

Online normalization also requires a window of data points to consider. There exist multiple windows models~\cite{zhu2003efficient} for online streaming data. Mainly, \begin{enumerate*}[label=(\roman*)]\item \textit{landmark windows}, which span from a landmark of the past to the present, \item \textit{sliding windows}, which have a fixed size and data points passing through them in a first-in-first-out fashion, \item \textit{damped windows}, which give data points weights decaying exponentially over time. \end{enumerate*}
We chose the damped window model due to its simple iterative calculation and the advantage of not requiring extra storage. The standardization parameters are set as exponentially weighted moving average (EWMA) and exponentially weighted moving variance (EWMV)~\cite{macgregor1993exponentially}, defined as follows:

\begin{algorithm}[!t]
\caption{SymED - Sender}\label{alg:SymED sender online compression}
\scriptsize
\begin{algorithmic}[1]
\Function{onlineCompression}{$tol, len_{max}$}
    \State get $T_s$ from memory
    \State $err \gets 0$; $bound \gets 0$; $len_{ts} \gets 0$
    \While{$err <= bound \textbf{ and } len_{ts} <= len_{max}$} \label{line:comp while}
        \State $t_j \gets \Call{getNextDataPoint}$
        \State append $t_j$ to $T_s$
        \State $\Call{updateOnlineNormalizationParams}{t_j}$ \label{line:update norm}
        \State $T_{sn} \gets$ standardize $T_s$ \label{line:use norm}
        \State $err \gets \Call{getError}{T_{sn}}$ \label{line:error}
        \State $len_{ts} \gets \Call{length}{T_s}$
        \State $bound \gets (len_{ts} - 2) * tol$ \label{line:bound}
    \EndWhile
    \State $T_s \gets$ last 2 elements of $T_s$
    \State store $T_s$ in memory
    \State $\Return$ first element of $T_s$
\EndFunction
\end{algorithmic}
\end{algorithm}

\begin{align}
       \label{eq:ewmat} EWMA_j = \alpha t_j + (1-\alpha)EWMA_{j-1} \\
        \label{eq:ewmvt} EWMV_j = \alpha(t_j-EWMA_j)^2 + (1-\alpha)EWMV_{j-1}
\end{align}
In Equation \ref{eq:ewmat} and \ref{eq:ewmvt}, $t_j$ indicates the next data point of the processed stream. 
The hyperparameter $\alpha$ serves as a weight, which has an exponentially decreasing influence on past data points. Here, EWMA and EWMV have same $\alpha$ value, for simplicity and consistency. Initially,  $EWMA_0 = t_0$ and $EWMV_0 = 1.0$ are set. All data kept in memory are standardized newly each iteration with up-to-date EWMA and EWMV. The update process of EWMA and EWMV, using Equations~\ref{eq:ewmat} and~\ref{eq:ewmvt}, can be found in Algorithm~\ref{alg:SymED sender online compression} (line~\ref{line:update norm}). Standardization is done through $\frac{t_h - EWMA_j}{\sqrt{EWMV_j}}$, e.g., for each data point $t_h$with $h \le j$ (line~\ref{line:use norm}).
\\
\indent \textbf{Online Compression: } In ABBA compression~\cite{elsworth2020abba}, data is approximated by a polygonal chain of linear pieces, where each piece is bounded by length and squared Euclidean distance error.
Linear pieces are defined as $P = [p_1, p_2,...,p_n]$, where each linear piece $p =(len, inc)$ is a tuple of length and increment value. Our proposed online compression Algorithm~\ref{alg:SymED sender online compression} only works on one linear piece at a time, instead of converting them all at once, like~\cite{elsworth2020abba}. After checking the error and maximal length limits in line~\ref{line:comp while}, one of the following cases can happen, \begin{enumerate*}[label=(\roman*)]\item no boundaries are reached and the algorithm continues the compression in the next iteration by trying to add another data point $t_{j+1}$ to the time series segment $T_s$, \item if $len_{max}$ is surpassed or the error including the current data point $t_j$ is out of bounds (see line~\ref{line:bound} for bound value), then the loop terminates \end{enumerate*}. After the loop, $T_s$ is set from $[t_0, ... ,t_m]$ back to the points $[t_{m-1}, t_m]$, to initialize the compression of the next segment. Finally, the endpoint of the segment $t_{m-1}$ is returned and sent to the receiver. Originally, the ABBA compression~\cite{elsworth2020abba} would use $T_s$ to produce a piece $p = (m-1, t_{m-1}-t_0)$ here, before moving on to compressing the next piece. However, in SymED, we move this step to the receiver. In this way, \begin{enumerate*}[label=(\roman*)]\item the size of payload needed to be transmitted is reduced by half, only sending one numeric value ($t_{m-1}$) instead of two ($p$), and \item making the receiver more robust to missing sender values.\end{enumerate*} 
Length and increment of a piece $p_i$ are always relative to its predecessor $p_{i-1}$. One missing piece would break up the polygonal chain of pieces ABBA depends on. SymED avoids this problem by only transmitting data points as absolute values from the sender to the receiver. 

Compressing $m$ data points to one linear piece with length $len = m-1$ requires $\mathcal{O}(m)$ iterations of the while loop (line~\ref{line:comp while}) and recalculating the error at line~\ref{line:error} in $\mathcal{O}(m)$ time, hence, Algorithm~\ref{alg:SymED sender online compression} runs in $\mathcal{O}(m^{2})$. For the whole data stream of size $N$, assuming each linear piece compresses on average $m$ data points ($m << N$), the complexity is $\mathcal{O}(N)$~\cite{elsworth2020abba}.

\subsection{ Receiver Side - Symbolic Conversion } 

\begin{algorithm}[t]
\caption{SymED - Receiver}\label{alg:SymED receiver}
\scriptsize
\begin{algorithmic}[1]
\Procedure{SymED}{$tol, scl, k_{min}, k_{max}$}
    \State $S \gets []$; $C \gets []$; $P \gets []$; $t_{i-1} \gets 0$
    
    \While{$True$}
        \State $t_i \gets \Call{getDataPointFromSender}$
        \State $len \gets \Call{timeSinceLastUpdate}$
        \State $inc \gets t_i - t_{i-1}$
        \State $p_i \gets (len, inc)$
        \State append $p_i$ to $P$
        \State $S , C \gets \Call{onlineDigitization}{P, C, tol, scl, k_{min}, k_{max}}$ \label{line:onlineDigi}
        \State $t_{i-1} \gets t_i$
    \EndWhile
\EndProcedure
\end{algorithmic}
\end{algorithm}

The job of the receiver is to listen for data points $t$ coming from sender devices and convert each of two subsequent data points to a linear piece $p$. All pieces $P$ are clustered in an online fashion, to get the converted sequence of symbols $S$, which essentially becomes one symbol longer after each received data point. Optionally, a reconstruction of the data stream can be done on demand. We decided to do the symbolic conversion at the receiver instead of the sender, because \begin{enumerate*}[label=(\roman*)]\item the sender is relieved of the computational demands, and \item symbolic conversion at the sender would require frequent and costly transmissions of the up-to-date reconstruction centers to the receiver\end{enumerate*}.

\indent \textbf{Construction of Linear Pieces: } The receiver Algorithm~\ref{alg:SymED receiver} receives data point $t_i$ in iteration $i$ from a sender. Along with data point $t_{i-1}$ of the previous iteration, the length and increment values $(len, inc)$ of the current linear piece $p_i$ can be constructed. We infer $len$ by taking advantage of the real-time online setting. To do that, the receiver saves timestamp $time_i$ upon the arrival of each $t_i$. Taking the difference in times with $len = time_i - time_{i-1}$ allows us not to have the sender transfer this value. Consequently, $inc = t_i - t_{i-1}$ completes the construction of $p_i$. Afterwards, at line~\ref{line:onlineDigi}, all pieces $P$ found so far get clustered to centers $C$ and converted to a symbolic string $S$ through Online Digitization in Algorithm~\ref{alg:SymED receiver online digitization}, which also determines the time complexity of Algorithm~\ref{alg:SymED receiver}.
\\ \indent \textbf{Online Digitization: } The Algorithm~\ref{alg:SymED receiver online digitization} uses clustering to group pieces $P = [p_1, p_2,...,p_n]$ to centers $C = [c_1, c_2,...,c_k]$. Each center $c$ represents a character of the alphabet $A = [a_1, a_2,...,a_k]$, mapping $P$ to the symbolic string $S = [s_1, s_2,...,s_n]$, and the center coordinates are responsible for the reconstruction of length and increment values of $P$. A scaling factor $scl$ is provided to weigh lengths of pieces differently from increments during 2D clustering, for $scl \in (0, \infty)$. The classical approach~\cite{elsworth2020abba} also considers that $scl \in \{0, \infty\}$, allowing for 1D-clustering either the lengths or increments, while $scl = 0$ is selected to put more emphasis on the trends of the time series. Our proposed SymED clustering can also be done either in 2D or in 1D, however, we focus mainly on 2D in this work. \\ \indent For SymED, we use a customized online version of k-means for both 1D and 2D clustering, because k-means is widely studied and provides a suitable streaming-based version~\cite{puschmann2016adaptive}, feasible for our online implementation. The steps of k-means, calculation of the cluster variances, and checking them against the tolerance boundary $tol^2_s$, follow the standard processes~\cite{elsworth2020abba}.%
\\ \indent In the online k-means function within Algorithm~\ref{alg:SymED receiver online digitization}, 
instead of the default initialization (randomized seeding), we initialize cluster centers $C_{init}$ with the values from the previous old clusters $C$, to remove the need for restarting a randomly initialized clustering~\cite{puschmann2016adaptive}. Consequently, the number of clusters $k$ for the first run of k-means is set to $k_o$, the number of old clusters in $C$, to avoid trying many values of $k$.  
If an additional cluster is still needed, $k$ is incremented by one, and the clustering is re-run (line~\ref{line:onlineKmeans while}). We initialize the newly added center with the newest piece, while the rest of the center initialization remains the same, ensuring fast convergence (line~\ref{line:append init}). Random-based initialization of centers is only chosen in line~\ref{line:rand init}, if the previous attempts of re-using old cluster centers fail. The $k_{min}$ and $k_{max}$ limit the number of clusters, as well as the size of the alphabet. After clustering is done, labels $L = [0, 1, ...]$, are mapped to symbols $[`a\textrm', `b\textrm', ...]$ and returned as string $S$, along with updated centers $C$.
\\ \indent The runtime of Algorithm~\ref{alg:SymED receiver online digitization} is bounded by the complexity of k-means. The average complexity to produce a new symbol is therefore $\mathcal{O}(kn)$ for $k$ clusters and $n$ linear pieces, per k-means iteration. Due to initialized centers and adding pieces one-by-one to the clusters, only very few iterations are needed. To convert a data stream of size $N$ to $n$ symbols, the resulting complexity is $\mathcal{O}(kn^2)$.

\begin{algorithm}[!t]
\caption{SymED - Receiver - Online Digitization}\label{alg:SymED receiver online digitization}
\scriptsize
\begin{algorithmic}[1]
\Function{onlineDigitization}{$P, C, tol, scl, k_{min}, k_{max}$}
    \If{$\Call{length}{C} < k_{min}$}
        \State $L \gets [0,1, ..., \Call{length}{P} - 1]$
        \State $S \gets \Call{labelsToSymbols}{L}$
        \State $C \gets P$
        \State $\Return$ $S, C$
    \EndIf
    \State standardize $P$ and $C$ and scale with $scl$
    \State $tol_s \gets \Call{getTolS}{tol, P}$; $len_{P} \gets \Call{length}{P}$
    \State $C_{init} \gets C$; $k_o \gets \Call{length}{C}$; $k \gets k_o - 1$;  $err \gets \infty$
    \While{$k < k_{max} \textbf{ and } k < len_{P} \textbf{ and } err > bound$} \label{line:onlineKmeans while}
        \State $k \gets k + 1$
        \If{$k = k_o + 1$}
            \State append last element of $P$ to $C_{init}$ \label{line:append init}
        \ElsIf{$k > k_o + 1$}
            \State randomly initialize $C_{init}$ \label{line:rand init}
        \EndIf
        \State $C, L \gets \Call{kmeans}{C_{init}, k}$
        \State $err \gets \Call{maxClusterVariance}{P, C, L, k}$
    \EndWhile
    \State de-standardize $P$, $C$ and de-scale with $scl$
    \State $S \gets \Call{labelsToSymbols}{L}$ \label{line:labelsToSymbols}
    \State $\Return$ $S, C$
\EndFunction
\end{algorithmic}
\end{algorithm}

\textbf{Reconstruction: } Converting a sequence of symbols $S$ back to a time series $\hat{T}$ follows three steps~\cite{elsworth2020abba}: 
\begin{enumerate*}[label=(\roman*)] \item \textit{Inverse-Digitization}, replacing $S$ with length and increment values $(\widetilde{len}, \widetilde{inc})$ of their corresponding reconstruction centers to reconstruct linear pieces, \item \textit{Quantization}, rounding lengths of those linear pieces back to whole numbers, generating $(\widehat{len}, \widehat{inc})$, and \item \textit{Inverse-Compression}, interpolating all-time series points for the chain of linear pieces, producing $\hat{T} = [\hat{t_0}, \hat{t_1}, ..., \hat{t_N]}$\end{enumerate*}. 
This offline reconstruction procedure from symbols works for both ABBA and SymED. Additionally, for SymED, a more accurate online reconstruction for $\hat{T}$ is possible by directly doing the Inverse-Compression step, with the original $(len, inc)$ values of pieces constructed by the receiver.

\section{Performance Evaluation}
\label{sec:evaluation}
\subsection{Experimental Setup}

\textbf{Metrics}: 
To measure the performance of SymED, we consider four main metrics. Namely, \begin{enumerate*}[label=(\roman*)]\item \textit{reconstruction error}, \item \textit{compression rate}, \item \textit{dimension reduction rate}, and \item \textit{computational latency} \end{enumerate*}. We measure reconstruction error ($RE$) through the Dynamic Time Warping (DTW) distance~\cite{berndt1994using}
between the original time series $T$ and the reconstruction $\hat{T}$, i.e., $RE = dtw(T, \hat{T})$, as in ~\cite{elsworth2020abba}. Additionally, for SymED, we evaluate the reconstruction error not only from symbols $S$, but also from linear pieces $P$, since they are also available for the SymED receiver. The compression rate ($CR$) for ABBA ($CR_{ABBA}$) and SymED ($CR_{SymED}$) is measured as defined in Equation~\ref{eq:CR}. Here, we measure how many bytes are saved during transmission from the sender to the receiver, instead of just sending an uncompressed raw data stream. We measure the dimension reduction rate ($DRR$), a measure of data size reduction while preserving the original data properties, by comparing lengths of converted symbols $S$ and true time series $T$, i.e., $DRR = \frac{len(S)}{len(T)}$. Here,  $len()$  returns the length of the input (count of symbols or data points). Dimension reduction helps to cope with the \textit{curse of dimensionality} when working with high-dimensional data.

\begin{align}
\label{eq:CR} CR_{ABBA} = \frac{bytes(C) + bytes(S)}{bytes(T)}   && CR_{SymED} = \frac{bytes(P) / 2}{bytes(T)}
\end{align}

In Equation \ref{eq:CR}, $bytes()$ returns a total number of bytes for the input. The assumptions of this experimental setting are, a symbol/character is a size of 1 byte, and a numerical/float value has a size of 4 bytes. $S$ is a series of symbols, $T$ is a series of floats, and $C$ is a set of centers, where each center is defined through 2 float valued coordinates. $P$ is a sequence of linear pieces, where a linear piece $p$ is defined over 2 float values. With ABBA, we assume the sender does the symbolic conversion offline in a batch, then sends all symbols $S$ and reconstruction centers $C$ to the receiver. For SymED, we only need to transmit one float value for each $p$, hence $bytes(P )/2$ for $CR_{SymED}$ in Equation~\ref{eq:CR}. For simplicity, any other bytes regarding a transmission protocol between the sender and receiver are omitted. For all metrics, a lower value means better performance.

\begin{table}[!b]
     \vspace{-8mm}

    \caption{Selected datasets of the UCR Time Series Classification Archive~\cite{UCRArchive2018}.}
    \centering
    \scriptsize
    \begin{tabular}{l|l|l|l}
    \hline
        \textbf{Dataset} & \textbf{Type} & \textbf{Size} & \textbf{Length} \\ \hline
        ACSF1 & Device & 10 & 1460 \\ 
        CinCECGTorso & Sensor & 4 & 1639 \\
        EOGHorizontalSignal & EOG & 12 & 1250 \\ 
        EOGVerticalSignal & EOG & 12 & 1250 \\ 
        EthanolLevel & Spectro & 4 & 1751 \\ 
        HandOutlines & Image & 2 & 2709 \\ 
        Haptics & Motion & 5 & 1092 \\ 
        HouseTwenty & Device & 2 & 2000 \\ 
        InlineSkate & Motion & 7 & 1882 \\ 
        Mallat & Simulated & 8 & 1024 \\ 
        MixedShapesRegularTrain & Image & 5 & 1024 \\ 
        MixedShapesSmallTrain & Image & 5 & 1024 \\ 
        PLAID & Device & 11 & 1344 \\ 
        Phoneme & Sensor & 39 & 1024 \\ 
        PigAirwayPressure & Hemodynamics & 52 & 2000 \\ 
        PigArtPressure & Hemodynamics & 52 & 2000 \\ 
        PigCVP & Hemodynamics & 52 & 2000 \\ 
        Rock & Spectrum & 4 & 2844 \\ 
        SemgHandGenderCh2 & Spectrum & 2 & 1500 \\ 
        SemgHandMovementCh2 & Spectrum & 6 & 1500 \\ 
        SemgHandSubjectCh2 & Spectrum & 5 & 1500 \\ 
        StarLightCurves & Sensor & 3 & 1024 \\ \hline
    \end{tabular}
    \vspace{-20mm}
    \label{table:dataset}
\end{table}

The final metric is computational latency, addressing the average amount of computational time needed for each symbol in the online setting. We measure the time required for a SymED sender to perform compression and a receiver to do symbolic conversion and reconstruction on a per-symbol basis. Compared to offline ABBA, we take the total time for all produced symbols, i.e., how long it takes on average to fully convert time series to symbols and reconstruct it again.
\\
\indent \textbf{Edge scenario setup:} We emulate the sender-receiver setup, where sender is an IoT sensor streaming pre-processed data towards receiver edge node for further processing. The setup is implemented as a multi-thread Python application.
SymED is split up as explained in Section~\ref{sec:conversion}.
For ABBA, we assume the sender does offline symbolic conversion of the time series and sends symbols and reconstruction centers to the receiver, where reconstruction happens. 
Evaluation is done on a Raspberry Pi 4B (4GB RAM).
\\
\indent \textbf{Datasets:} We use UCR Time Series Classification Archive~\cite{UCRArchive2018} datasets as a representative of IoT 
data~\cite{elsworth2020abba}. We filter the \textit{test} split for datasets with a minimal length of 1000 data points, ensuring we have sufficient data for the online normalization to adapt.
We sample each dataset by selecting the first time series of each class, e.g., for dataset \textit{ACSF1} with a size of 100 time series and 10 different classes, we take a sample of 10 time series, each with a length of 1460.
Table~\ref{table:dataset} shows 22 selected datasets containing 302 time series with mean length of 1673. 
\\
\indent \textbf{ Baseline and Hyperparameters:} We compare the results of our proposed SymED to the original ABBA, a baseline for reconstruction accuracy. Compared to ABBA, SymED has an additional hyperparameter $\alpha$ for adjusting the weights of  online normalization values EWMA and EWMV. Higher $\alpha$ values prefer the most recent data, monitoring short-term variability of EWMA and EWMV, and lower values focus on long-term estimation of mean and variance~\cite{macgregor1993exponentially}. We set  $0.01 \le \alpha \le 0.02$ based on empirical testing, suiting our chosen datasets. Further, we set $k_{min} = 3$ for both ABBA and SymED, meaning that an alphabet of at least three symbols will be used. The only exception is when $|P| < k_{min}$, where too few linear pieces are in $P$ to form $k_{min}$ clusters, resulting $k_{min} = |P|$. We set $k_{max} = 100$, the upper bound for the alphabet size.
\\
For each algorithm and tolerance value, the mean of the results over all datasets (Table~\ref{table:dataset}) is taken. To compensate for the different sizes of datasets, we assign equal weights in the evaluation, i.e., averaging results first for all time series within a dataset, then taking the average once again over all datasets.

\begin{figure*}[tb]
    \centering
    \begin{subfigure}[t]{0.2135\textwidth}
        \centering
        \includegraphics[width=\textwidth]{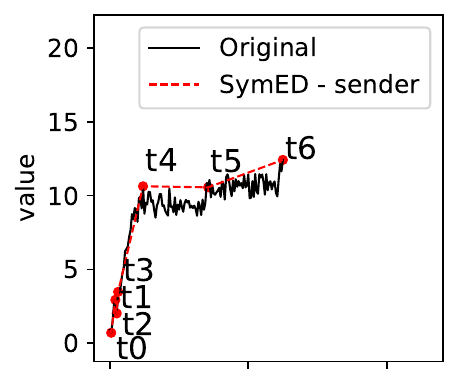}
        \captionsetup{skip=-1pt}
        \caption{}
        \label{fig:runn-a}
    \end{subfigure}
    \begin{subfigure}[t]{0.18\textwidth}
        \centering
        \includegraphics[width=\textwidth]{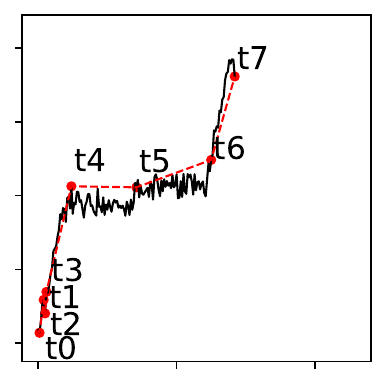}
        \captionsetup{skip=-1pt}
        \caption{}
        \label{fig:runn-b}
    \end{subfigure}
    \begin{subfigure}[t]{0.18\textwidth}
        \centering
        \includegraphics[width=\textwidth]{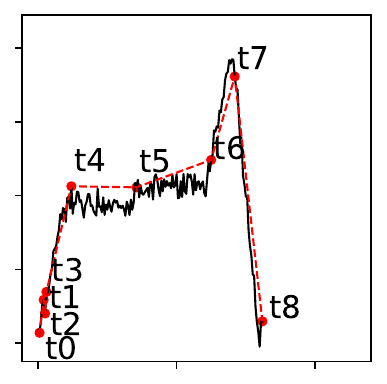}
        \captionsetup{skip=-1pt}
        \caption{}
        \label{fig:runn-c}
    \end{subfigure}
    \begin{subfigure}[t]{0.18\textwidth}
        \centering
        \includegraphics[width=\textwidth]{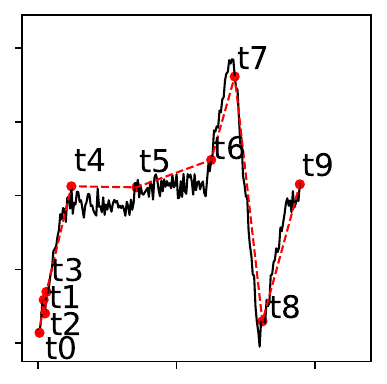}
        \captionsetup{skip=-1pt}
        \caption{}
        \label{fig:runn-d}
    \end{subfigure}
    \begin{subfigure}[t]{0.18\textwidth}
        \centering
        \includegraphics[width=\textwidth]{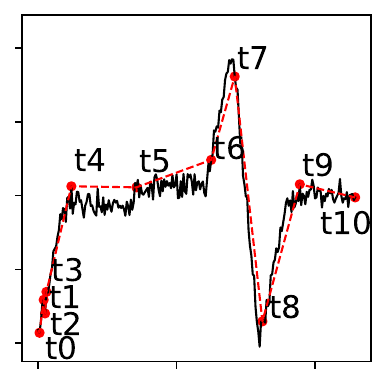}
        \captionsetup{skip=-1pt}
        \caption{}
        \label{fig:runn-e}
    \end{subfigure}

    \centering
    \begin{subfigure}[t]{0.2135\textwidth}
        \centering
        \includegraphics[width=\textwidth]{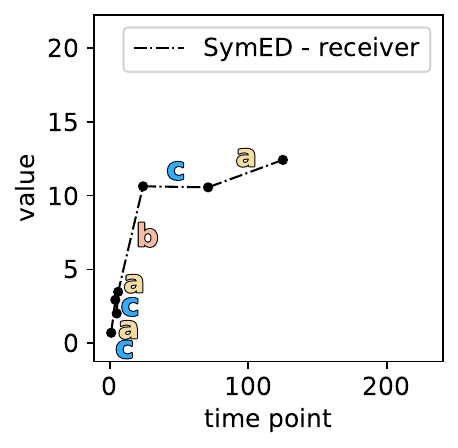}
        \captionsetup{skip=-1pt}
        \caption{}
        \label{fig:runn-f}
    \end{subfigure}
    \begin{subfigure}[t]{0.18\textwidth}
        \centering
        \includegraphics[width=\textwidth]{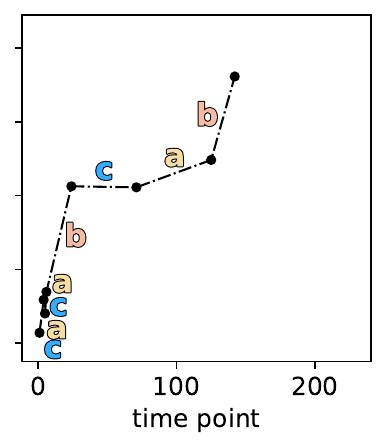}
        \captionsetup{skip=-1pt}
        \caption{}
        \label{fig:runn-g}
    \end{subfigure}
    \begin{subfigure}[t]{0.18\textwidth}
        \centering
        \includegraphics[width=\textwidth]{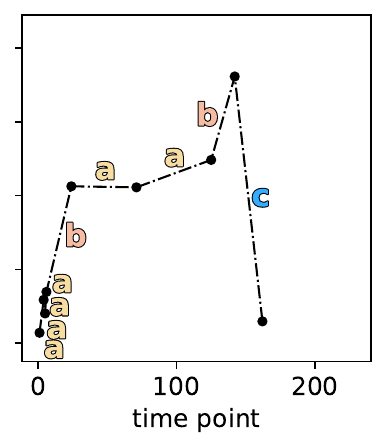}
        \captionsetup{skip=-1pt}
        \caption{}
        \label{fig:runn-h}
    \end{subfigure}
    \begin{subfigure}[t]{0.18\textwidth}
        \centering
        \includegraphics[width=\textwidth]{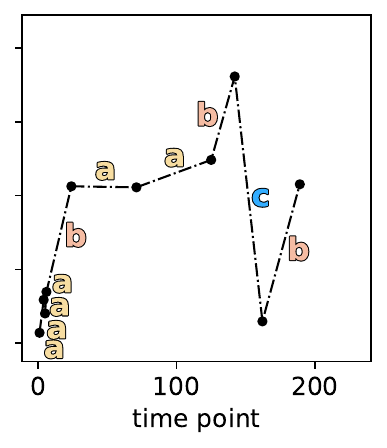}
        \captionsetup{skip=-1pt}
        \caption{}
        \label{fig:runn-i}
    \end{subfigure}
    \begin{subfigure}[t]{0.18\textwidth}
        \centering
        \includegraphics[width=\textwidth]{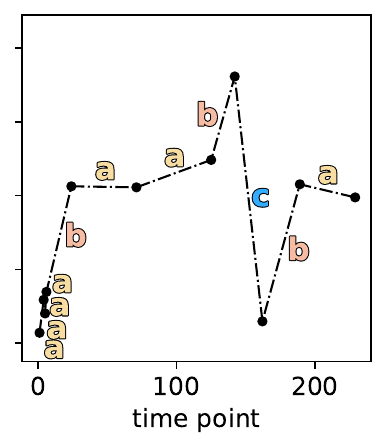}
        \captionsetup{skip=-1pt}
        \caption{}
        \label{fig:runn-j}
    \end{subfigure}
    \caption{Running example for SymED sender (a-e) and receiver (f-j) algorithms.}
    \label{fig:running example}
\end{figure*}

\subsection{Running Example}
We provide a running example in Figure~\ref{fig:running example}, on a time series example of 230 data points, similar to the ABBA~\cite{elsworth2020abba}. Here, parameters are set as $tol = 0.4$, $\alpha = 0.02$, and $scl = 0$ (1D clustering). In Figure~\ref{fig:running example}, the sender-side (IoT nodes) steps are depicted in Figures~\ref{fig:runn-a}-\ref{fig:runn-e}, the receiver side steps (edge nodes) in Figures~\ref{fig:runn-f}-\ref{fig:runn-j}. Each sub-figure shows the generation of one new linear piece and symbol, from left to right. For brevity, we summarized the first seven iterations in Figures~\ref{fig:runn-a} and ~\ref{fig:runn-f}, then showed the remaining iterations in the remaining figures.

\par The sender compresses the incoming data stream (solid black line) until a linear piece $p_i$ is formed (red dashed line) and then sends the endpoint $t_i$ of $p_i$ to the receiver. The receiver reconstructs $p_i$ (black dash-dotted line) from $t_i$, and $t_{i-1}$, and does an online clustering to produce the symbol $s_i$ (`a', `b', or `c' here). SymED produces 11 symbols in total, namely, $aaaabaabcba$. 
At the beginning, the first four symbols are produced in very short intervals, due to the online normalization not having adapted to the data yet and also capturing noise. But afterwards, longer linear pieces start to get formed to produce the remaining symbols.
Due to the nature of online clustering, older pieces may be assigned to a different cluster after several updates.
This can be seen for a linear piece between $t_4$ and $t_5$, which changes from `c' to `a' (from Figure~\ref{fig:runn-g} to Figure~\ref{fig:runn-h}).

\subsection{Results and Analysis}

\begin{figure*}[tb]
    \centering
    \captionsetup[subfigure]{justification=justified,singlelinecheck=true,margin={0cm,11.5cm}}
    \begin{subfigure}[t]{\textwidth}
        \centering
        \includegraphics[width=0.9\textwidth]{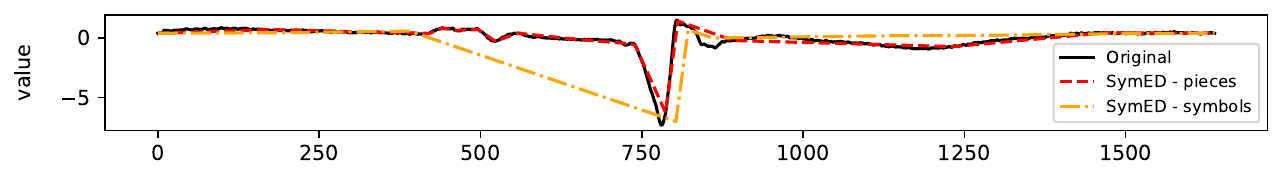}
        \captionsetup{skip=-39pt}
        \caption{}
        \label{fig:dataset example a}
    \end{subfigure}
    \begin{subfigure}[t]{\textwidth}
        \centering
        \includegraphics[width=0.9\textwidth]{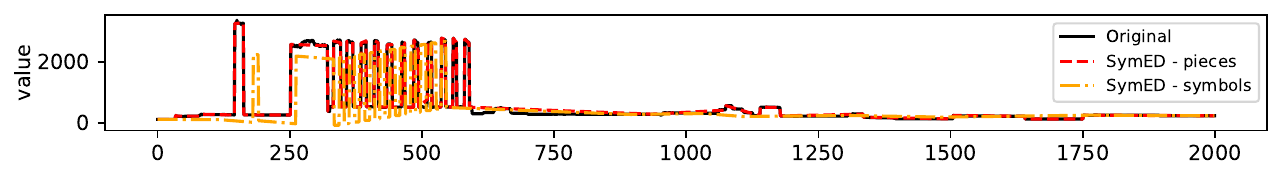}
        \captionsetup{skip=-39pt}
        \caption{}
        \label{fig:dataset example b}
    \end{subfigure}
    \begin{subfigure}[t]{\textwidth}
        \centering
        \includegraphics[width=0.9\textwidth]{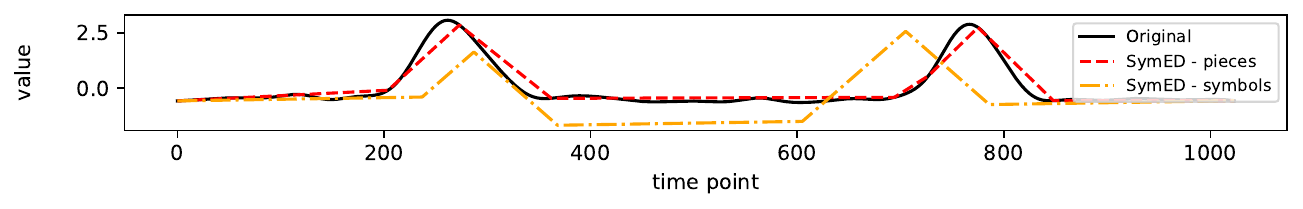}
        \captionsetup{skip=-45pt}
        \caption{}
        \label{fig:dataset example c}
    \end{subfigure}
    \caption{SymED reconstruction example on three representational datasets from the UCR archive: (a) \textit{CinCECGTorso}, (b) \textit{HouseTwenty}, (c) \textit{StarLightCurves}}
    \label{fig:dataset example}
\end{figure*}

\begin{figure*}[tb]
    \centering
    \begin{subfigure}[t]{0.32\textwidth}
        \centering
        \includegraphics[width=\textwidth]{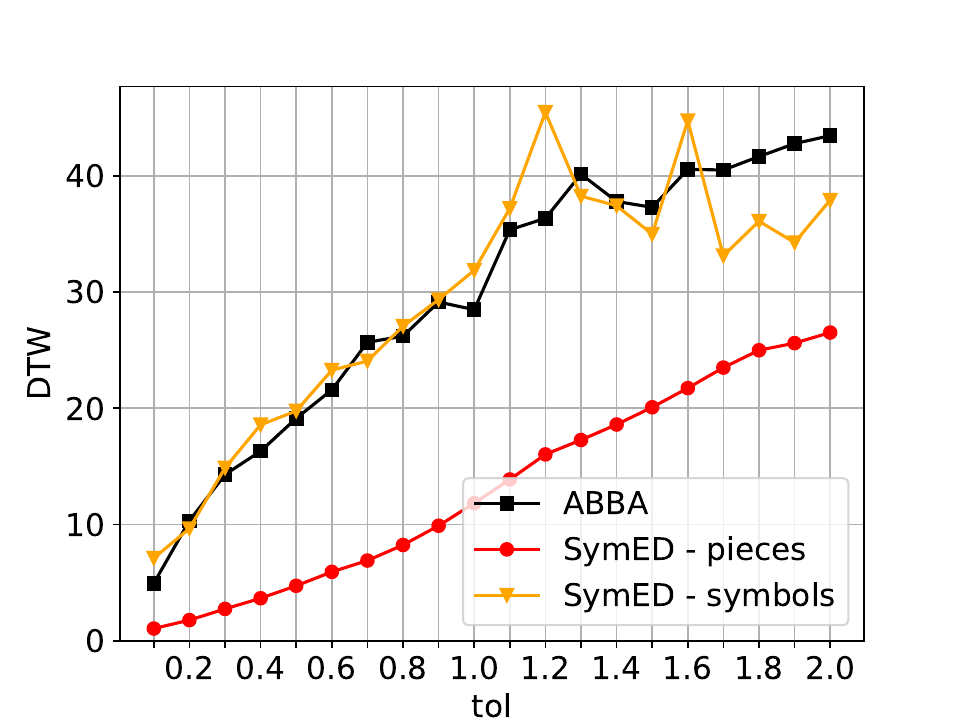}
        \captionsetup{width=.9\linewidth}
        \caption{Reconstruction Error }
        \label{fig:RE}
    \end{subfigure}
    \begin{subfigure}[t]{0.32\textwidth}
        \centering
        \includegraphics[width=\textwidth]{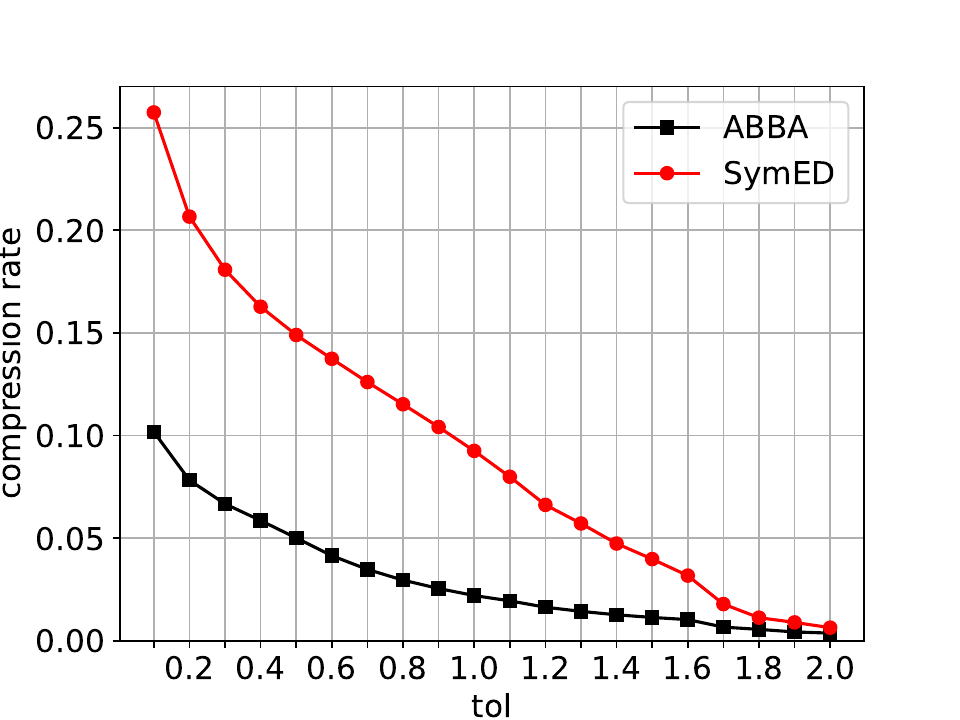}
        \captionsetup{width=.9\linewidth}
        \caption{Compression Rate }
        \label{fig:CR}
    \end{subfigure}
    \begin{subfigure}[t]{0.32\textwidth}
        \centering
        \includegraphics[width=\textwidth]{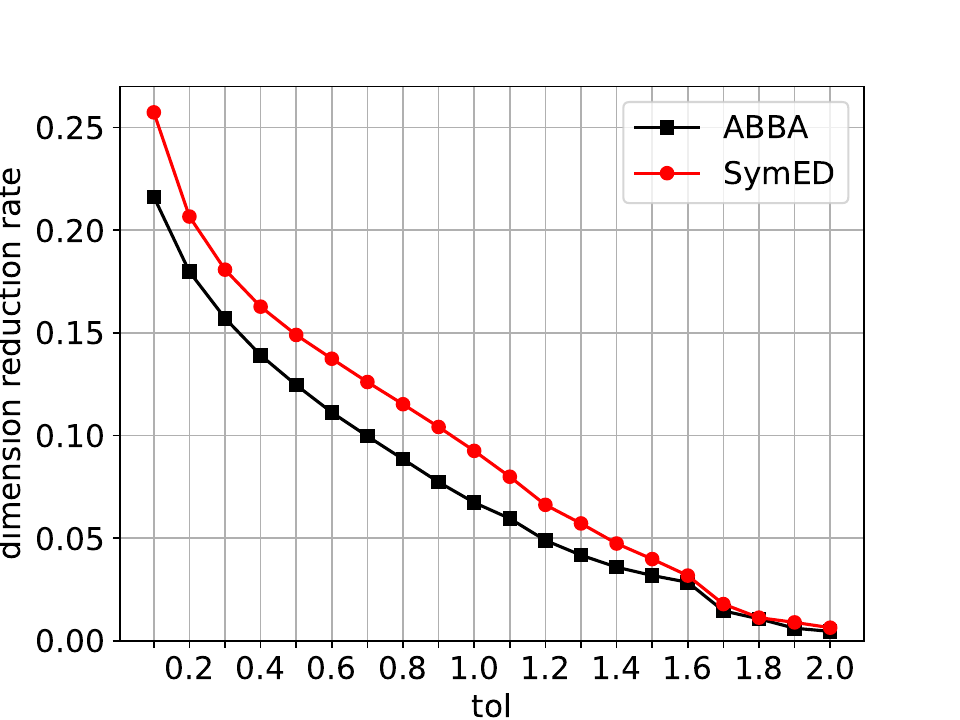}
        \captionsetup{width=.9\linewidth}
        \caption{Dim. Reduction Rate}
        \label{fig:DRR}
    \end{subfigure}
    
    \centering
    \begin{subfigure}[t]{0.32\textwidth}
        \centering
        \includegraphics[width=\textwidth]{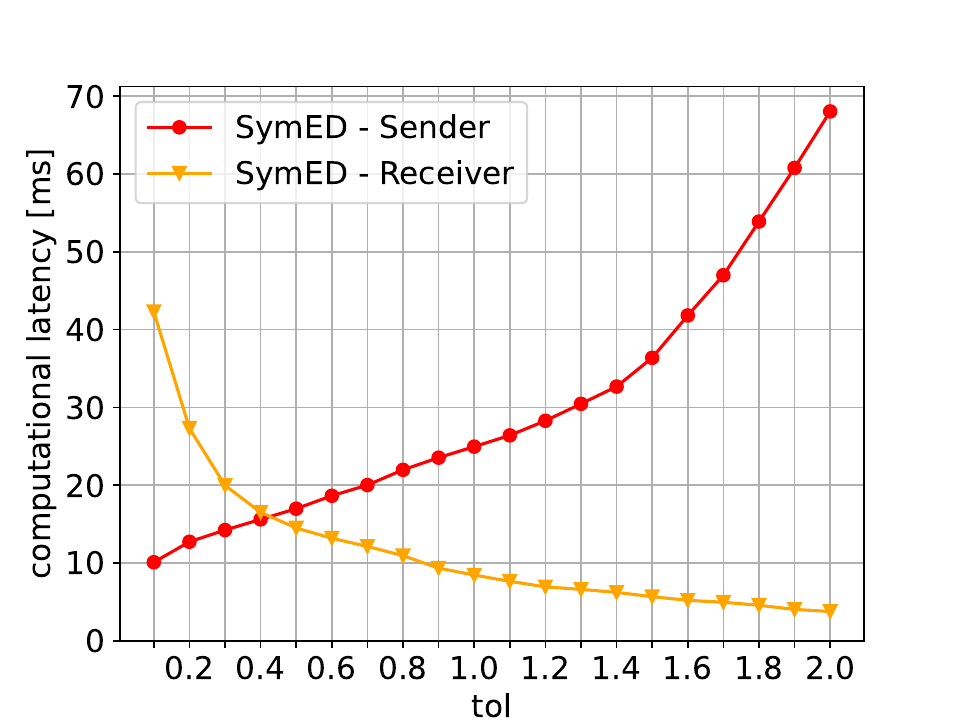}
        \captionsetup{width=.9\linewidth}
        \caption{%
        Online Latency}
        \label{fig:CL-SR}
    \end{subfigure}
    \begin{subfigure}[t]{0.32\textwidth}
        \centering
        \captionsetup{width=.9\linewidth}
        \includegraphics[width=\textwidth]{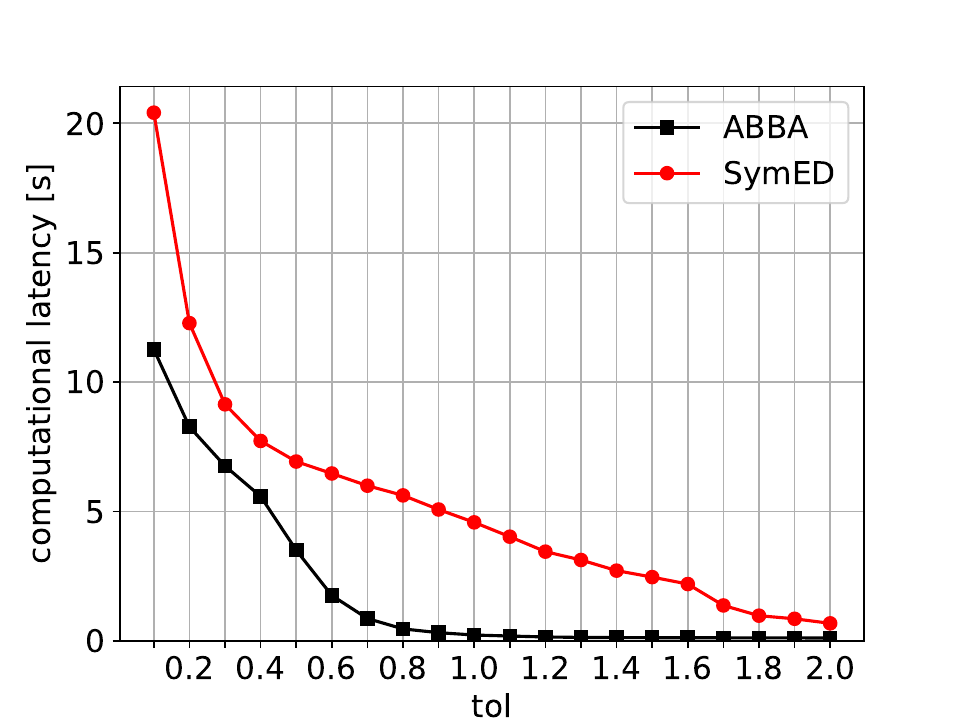}
        \caption{%
        Offline Latency}
        \label{fig:CL-T}
    \end{subfigure}
  
    \caption{Evaluation of ABBA and proposed SymED (averaged over all datasets).}
    \label{fig:eval}

\end{figure*}

Figures~\ref{fig:dataset example a}-\ref{fig:dataset example c} show examples of SymED reconstruction on a few UCR time series, using tolerance $tol=0.4$. The following metrics in Figure~\ref{fig:eval} are evaluated for a range of $tol$ values, going from $0.1$ to $2.0$ in $0.1$ increments. Other common parameters for Figures~\ref{fig:dataset example}-\ref{fig:eval} are $\alpha = 0.01$ and $scl=1.0$, using 2D clustering.
\\
\indent \textbf{Reconstruction error:} Figure~\ref{fig:RE} shows that SymED reconstruction error for symbol generation follows the original ABBA curve, which is a desired behavior. Reconstruction errors from symbols average around 29.25 for SymED and 29.60 for ABBA. In contrast, SymED online reconstruction from linear pieces has less than half the error at 13.25, due to pieces being more true to the original data, before being clustered and converted to symbols.
\\
\indent \textbf{Compression rate:} We compare the results of compression rates in Figure~\ref{fig:CR}, measuring the size reduction of transmitted data.
As seen in Figure~\ref{fig:CR}, ABBA compresses data to 3.1\% on average, by taking advantage of transmitting already converted symbols, which are less byte expensive than numerical data points of SymED. SymED's online and distributed nature comes at the cost of having a worse compression rate of 9.5\% on average.
\\
\indent \textbf{Dimension reduction rate:} 
Figure~\ref{fig:DRR} shows dimension reduction results. Both ABBA and SymED have similar behavior, since their compression phases work in a similar way. Differences occur due to the online normalization of SymED, which takes time to adapt to the data and produces a higher number of linear pieces/symbols early on, also evidenced in Figure~\ref{fig:running example}.
Finally, the SymED has a mean dimension reduction rate of 9.5\%, ABBA averages at 7.7\%.
\\
\indent \textbf{Computational latency:} Figure~\ref{fig:CL-SR} compares SymED sender and receiver, how long processing takes per symbol. Lower tolerances produce many short pieces, making clustering at the receiver dominant. In contrast, higher tolerance values produce fewer and longer linear pieces, increasing the compression times for the sender.
On average, a SymED sender spends 30ms on compressing, and a receiver 12ms on creating and reconstructing a symbol, summing up to 42ms total per symbol.
In Figure~\ref{fig:CL-T} we show the total latencies for processing an entire time series offline.
ABBA is overall faster with a mean of 2.0s, compared to 5.3s for SmyED,  however, SymED is mainly designed for online processing.

To conclude, SymED provides the benefit of lower online reconstruction error and real-time adaptability to streaming data, with a little  cost on higher data transmission needs and computational times compared to offline ABBA.

\section{Related Work}
\label{sec:related}
Symbolic representation (SR) algorithms have been used to convert time series data into symbols. The basic algorithm in the symbolic conversion is SAX~\cite{lin2007experiencing}. Another variant of SAX is proposed in \cite{liu2021online}, dedicated to online load data compression and reconstruction. The authors split the time series into the event and steady-state segments, while using symbolic conversion only on the latter one. In this version, the alphabet is fixed, while the window length is adaptive, by dividing segments into windows of equal information content. Although they use adaptive window sizes, as in our proposed SymED, they focus on event-based data instead of arbitrary time series.
In \cite{khan2014novel}, the author converts sensor data streams to symbols using SAX, followed by classification with a Support Vector Machine (SVM). Works like~\cite{kolozali2014knowledge} symbolize sensor data streams using SAX and incorporate data stream annotation in a distributed environment, interacting over a publish/subscribe messaging service. Further, SensorSAX \cite{ganz2013information}, is a SAX variation with dynamic window length, to reduce the energy consumption of IoT sensor streams. While using symbolic conversion to process IoT data, other works lack adaptability by either using a static window size~\cite{khan2014novel,kolozali2014knowledge} or fixed alphabet~\cite{khan2014novel,kolozali2014knowledge,ganz2013information}. They also sample the data stream and produce symbols in batches, in contrast to producing symbols consecutively in SymED.
Adaptive compression of IoT data based on different resource-limited edge conditions is proposed by~\cite{lu2020adaptively}. However, only the impact on edge-cloud bandwidth and data transfer is considered, without addressing the impact on edge analytics. \cite{puschmann2016adaptive} targets an adaptive streaming-based version of k-means. This solution starts with initial candidate clusters, trying to assign each new data point in the online phase to the nearest cluster, and only does a complete re-clustering if the clusters are not valid anymore. A validity check is done by analyzing the input stream's probability density function, where high deviations signal a concept data drift and require a new cluster initialization.
Still, they do not consider the tolerance-dependent variance checks of clusters, as in SymED. Similarly, \cite{trivedi2020sharing} considers data-sharing edge concepts, while \cite{azar2019energy} deals with the bandwidth limitation. However, no online concepts are considered with IoT data streams.
\\ \indent Although there exist different techniques for raw data compression in cloud and edge \cite{wang2020joint}, we particularly focus on SR for the edge. 
SR allows for direct analytics on compressed data, while also enabling reconstruction of the original data.
We believe this is a crucial advantage over other raw data compression techniques, reducing both network and storage usage for critical IoT systems.

\section{Conclusions and Future Work} \label{sec:conclusion}
We proposed SymED, a real-time online symbolic representation method for resource-constrained edge environments. We distribute the symbolic conversion workload between IoT sender and edge receiver devices, and also minimize the number of transmitted bytes between them. 
Hyperparameters in SymED, such as $tol$, balance reconstruction error and compression performance, while $\alpha$ determines the adaptability to streaming data through online normalization.
SymED achieves on average 9.5\% on compression rate and dimension reduction rate, with a mean online reconstruction error of 13.25 in the DTW space, while taking a mean time of 42ms to compute a symbol. 
Online SymED improves on reconstruction accuracy and adapting to data stream distribution, with a slight overhead in compression and computational efficiency, compared to the offline base algorithm ABBA. Our future plans involve enhancing SymED's performance for time-critical IoT applications by incorporating different clustering mechanisms.

\section*{Acknowledgements and Data Availability}
This work has been partially funded through the Runtime Control in Multi Clouds (RUCON), Austrian Science Fund (FWF): Y904-N31 START-Programm, 2015,
Sustainable Watershed Management Through IoT-Driven Artificial Intelligence (SWAIN), CHIST-ERA-19-CES-005, Austrian Science Fund (FWF), 2021,
Standalone Project Transprecise Edge Computing (Triton), Austrian Science Fund (FWF): P 36870-N, 2023,
Flagship Project High-Performance Integrated Quantum Computing (HPQC) \# 897481 Austrian Research Promotion Agency (FFG), 2023.
The artifact associated with this paper is available in the \textit{figshare} repository~\url{https://doi.org/10.6084/m9.figshare.23536992}. 

 \bibliographystyle{splncs04}
 \bibliography{reflist}

\end{document}